\documentclass{article}

\usepackage[preprint]{neurips_2025}

\usepackage[utf8]{inputenc} 
\usepackage[T1]{fontenc}    
\usepackage{hyperref}       
\usepackage{url}            
\usepackage{booktabs}       
\usepackage{amsfonts}       
\usepackage{nicefrac}       
\usepackage{microtype}      
\usepackage{xcolor}         
\usepackage{graphicx}
\usepackage[title]{appendix}
\usepackage{tabularx}
\usepackage{amsmath}

\title{From Black Box to Biomarker: Sparse Autoencoders for Interpreting Speech Models of Parkinson’s Disease}

\author{%
Peter Plantinga$^{1, 2, 3}$
Jen-Kai Chen$^{1, 2}$
Roozbeh Sattari$^{2, 4}$
Mirco Ravanelli$^{2, 3, 5}$
Denise Klein$^{1, 2}$ \\
$^1$ McGill University%
\qquad $^2$ Centre for Research on Brain, Language, and Music \\
$^3$ Mila - Quebec AI Institute%
\qquad $^4$ Douglas Research Centre%
\qquad $^5$ Concordia University%
}

\begin{document}

\maketitle

\begin{abstract}
  Speech holds promise as a cost-effective and non-invasive biomarker for neurological conditions such as Parkinson’s disease (PD). While deep learning systems trained on raw audio can find subtle signals not available from hand-crafted features, their black-box nature hinders clinical adoption. To address this, we apply sparse autoencoders (SAEs) to uncover interpretable internal representations from a speech-based PD detection system. We introduce a novel mask-based activation for adapting SAEs to small biomedical datasets, creating sparse disentangled dictionary representations. These dictionary entries are found to have strong associations with characteristic articulatory deficits in PD speech, such as reduced spectral flux and increased spectral flatness in the low-energy regions highlighted by the model attention. We further show that the spectral flux is related to volumetric measurements of the putamen from MRI scans, demonstrating the potential of SAEs to reveal clinically relevant biomarkers for disease monitoring and diagnosis.
\end{abstract}

\section{Introduction}
\label{sec:introduction}

Recorded speech is an attractive candidate for use as a biomarker, as it is inexpensive and straightforward to collect, and it reflects both motor and cognitive functions. This makes it particularly valuable for monitoring neurodegenerative diseases such as Parkinson's disease which are known to affect both motor and cognitive areas of the brain, through changes in articulation, prosody, and fluency (\cite{godino2017towards}). While deep-learning systems trained on raw audio have shown promising accuracy in detecting disease-related changes (\cite{favaro2024unveiling}), a lack of understanding about how these models arrive at a given prediction poses a major barrier to clinical adoption (\cite{ramanarayanan2022speech}).

Two primary strategies have been pursued so far to try and explain the behavior of deep learning models in speech: (1) building interpretable-by-design systems which incorporate mechanisms for explanation into the \emph{model structure and training} directly, and (2) generating post-hoc interpretations which analyze \emph{model behavior} after training is complete.

The ``interpretable-by-design'' strategy can work in a few different ways. Prototype networks operate by developing a representative example for each class and then for new inputs they find the closest example in order to produce a prediction (\cite{zinemanas2021interpretable, ren2022prototype}). However, the relevant examples may not completely explain a prediction as there may be multiple similar aspects and this technique does not explain which is responsible for a given prediction. An alternative approach attempts to build high-level concepts into the model itself. One example of this is called concept bottleneck networks (\cite{forest2024interpretable}). However, this approach requires foreknowledge of the relevant concepts, and it may be challenging to select a complete set of concepts, as they may not be known (\cite{rombach2021contrastive}).

By contrast, ``post-hoc explanation'' strategies typically operate by localizing decision causes to specific relevant features or signal regions (\cite{parekh2022listen,paissan2024listenable}). However, the typical localization techniques for audio can result in attributions that are difficult to interpret for biomarker-related tasks, as they are noisy and scattered throughout the spectral signal (\cite{Mancini2024ParkinsonSpeechExplainability}). This is the case for global attributes, such as breathy or quiet speech which are present throughout a signal. For some data there is a degree of localization, but the localization itself is not sufficient to explain the network behavior, as it is not clear what particular features of the region of interest are being attended. One example could be attending to regions of pauses between speech segments -- are the duration or number or ratio of pauses more relevant, or the sounds occurring within the pause?

To build a bridge between these approaches, we have adapted \emph{sparse autoencoders} (SAEs) (\cite{bricken2023towards}) from large language model (LLM) mechanistic interpretability research for the domain of speech-based PD detection. SAEs learn a sparse dictionary to approximate internal model activations, aiming to disentangle highly distributed representations into a small number of interpretable components. This method draws on the theory that sparse coding promotes more meaningful and monosemantic representations (\cite{olshausen1997sparse}). This method does not require any changes to the model training procedure, but can still make use of expert knowledge about similar examples and relevant high-level concepts.

There are two main difficulties in applying the SAE technique to speech biomarker problems:

\textbf{Dataset size} -- Biomedical speech datasets tend to be privacy-restricted and the relevant populations tend to make up only a fraction of the total population, reducing the feasibility of collecting large, highly-expressive dictionaries as done in LLM research. This limits the expressiveness of the dictionary, meaning smaller overall dictionaries and fewer activations.

\textbf{Dictionary interpretation} -- Unlike text, which is inherently symbolic, the audio domain is continuous and context-dependent, making it harder to associate meaning with co-occurrence patterns. With text it is possible to manually examine the instances that activate for a particular dictionary entry and look for related words or concepts, but for audio the plethora of potential connections between samples make a similar analysis more challenging.

In order to contend with these difficulties, we have developed novel strategies for applying SAEs to speech biomarker problems. To summarize our contributions:

\begin{itemize}
    \item We propose a novel masking structure, a single application of SAE per sample, and small SAE dictionary sizes to enable SAE training on biomedical speech data.
    \item We show that for PD detection, SAEs correlate with hand-crafted acoustic features, revealing that the model makes a prediction largely based on spectral flux and flatness in low-energy regions of the audio.
    \item We further demonstrate that the spectral flux is associated with volumetric measures of putamen size from magnetic resonance imaging (MRI) brain scans, linking speech signs directly with the underlying neuroanatomy.
\end{itemize}

These findings suggest that sparse autoencoders can serve as an effective tool for explaining the behavior of speech-based deep learning models, creating more trustworthy systems for future clinical use of machine learning in diagnosing and monitoring neurodegenerative diseases.


\begin{figure}
    \centering
    \includegraphics[width=0.9\linewidth]{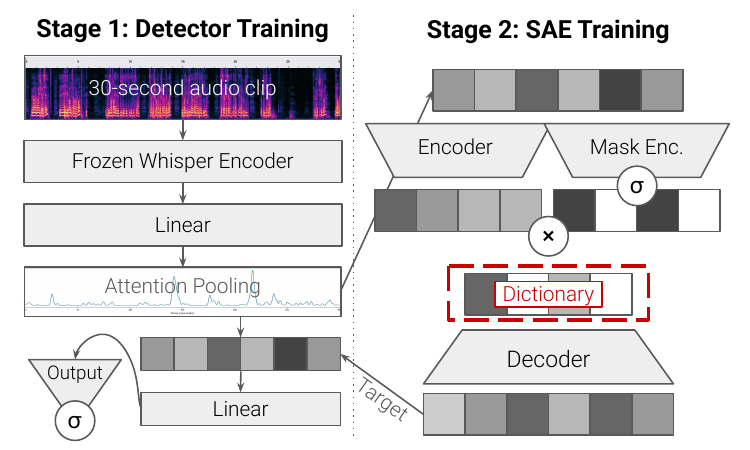}
    \caption{Overall Mask SAE system design. In the first stage, the detection parameters (left side) are trained, then frozen. In the second stage, the SAE (right side) is trained.}
    \label{fig:mask_system}
\end{figure}

\section{Related work}
\label{sec:related}

Research in the detection of Parkinson's disease from speech has long focused on interpretable measures based on acoustic signals. One common task is sustained vowel phonation (SVP), which started as early as 1963 (\cite{brown1963organic}) and continues through today (\cite{ali2024parkinson}). Another phonation task used for detection is the computation of the Vowel Space Area (VSA) and Vowel Articulation Index (VAI) that measures formant frequency spread for different vowels, which is reduced in PD patients (\cite{skodda2011vowel}). However, there is some research that disputes the effectiveness of vowel-based measures (\cite{skodda2012impairment, rusz2013imprecise}). There may still be a way forward for phonation tasks through the application of deep learning, as \cite{rahmatallah2025pre} and others have found.

Continuous speech may provide a better basis for disease detection (\cite{quan2021deep}), and again deep learning systems prove to be more predictive than interpretable features (\cite{favaro2024unveiling}). Recent attention has started to accrue to the idea that speech foundation models on continuous speech can contribute to a more straightforward and generalizable method for detecting PD (\cite{ali2024parkinson}). Some have found that these models can work in real-world acoustic conditions with the help of voice activity detection, speech enhancement, and dereverberation models (\cite{la2024exploiting}). Other work has explored strategies for improving generalizability across languages using multiple tasks (\cite{laquatra2025bilingual}), as well as improving the aggregation of dysarthric cues across speech segments with graph neural networks (\cite{sheikh2025graph}). However, interpretation is hard for this domain, and traditional interpretability techniques do not work well (\cite{Mancini2024ParkinsonSpeechExplainability}).

The failure of traditional techniques has led us to look to other fields for interpretability techniques. Sparse autoencoders were designed for interpretation of large lancome from the field of mechanistic interpretability, where there have been several developments since the introduction of SAEs for interpretation. \cite{rajamanoharan2024jumping} introduce the use of JumpReLU activations and an L0 sparsity objective in order to avoid influencing the scale of the results, which can happen with the use of L1 as a sparsity objective. For better control of the number of activations per sample, TopK was introduced by \cite{gao2024scaling} and improved with BatchTopK (\cite{bussmann2024batchtopk}). Unlike these approaches, our smaller datasets and audio-based inputs mean we need more flexibility in the generation of our dictionary.

Finally, there are some works that seek to achieve interpretable systems with deep learning for Parkinson's disease detection from speech, such as the work of \cite{gimeno2025unveiling}. In this work the authors use a fusion of interpretable and deep learning features to explain system behavior. However, the presence of interpretable features by themselves don't explain the behavior of the deep learning portion of the network. By contrast, \cite{shah2022parkinson} follow a post-hoc localization approach, dividing the audio into logical chunks and ranking the chunks in terms of impact on the final decision. While localization is important, our work also investigates high-level attributes of the decision, such as what particular qualities of the audio were attended to.

\section{Methods}
\label{sec:methods}

The overall system for interpretable biomarker design is in two stages, which can be seen in the two halves of Figure \ref{fig:mask_system}. Each stage is described in more detail below.

\subsection{Parkinson's disease detection from speech}
\label{ssec:detection}

In the first stage, we built a system for accurate detection by leveraging frozen speech foundation models for detection and adding a minimal set of aggregation parameters to produce a prediction for each speech sample. We used an attention pooling layer to combine the sequence of embeddings into a fixed-dimensional vector for prediction (\cite{safari2020self}). We trained the detection parameters using a binary cross-entropy objective against the binary PD / HC label.

To verify that this basic system is working well, we tested a variety of encoders and chose to apply SAEs to Whisper Small (\cite{radford2022whisper}), as it is one of the two top-performing systems, while remaining more computationally efficient than the best system, which is Whisper Medium. The full list of results can be seen in Appendix \ref{apx:detection}.

In addition to pretrained models, the appendix includes results using Mel-scale filterbank spectral features and extracted vocal quality features, which serve as baselines for comparing the pretrained model features. The vocal features are a combination of autocorrelation-based features (F0, harmonicity, glottal-to-noise-excitation ratio~\cite{godinollorente2010gne}), period-based features (jitter, shimmer), spectral-based features (e.g. skew, flux), and the first 4 mel-frequency cepstral coefficients. 

It is important to note here that the same exact features that later provide a strong interpretation of the Whisper-based detection model behavior do not by themselves constitute a basis for strong disease prediction. This could be due to a better localization of the regions of interest by Whisper. Perhaps Whisper is using language-related features that are inaccessible through the raw acoustic features.

\subsection{Sparse autoencoder for interpretation}
\label{ssec:method_sae}

After fine-tuning for detection, we froze all detection parameters and inserted a SAE layer immediately after the attention pooling layer. This layer produces a small dictionary where each entry corresponds to a meaningful aspect of the final decision, while remaining sufficiently upstream of the decision as to represent important aspects of the original input.

Let $x \in \mathbb{R}^N$ be the dense output of the attention pooling. The SAE encodes $x$ into a sparse dictionary representation $f(x) \in \mathbb{R}^K$ via dual encoder projections: $W_e, W_m \in \mathbb{R}^{K \times N}$, to separate the computation of each entry value from the decision to activate or inactivate the entry. Each projection also contains a bias term $b_e, b_m \in \mathbb{R}^K$. We use a temperature-controlled sigmoid for the mask via a temperature parameter $\tau$ which is annealed from 1.0 to 0.2 over the first 200 training steps.

\begin{align}
    \text{mask}(x) &= \sigma(\tau \cdot (W_m x + b_m)) \\
    f(x) &= (W_e x + b_e) \odot \text{mask}(x)
\end{align}

To produce a reconstruction of the original vector $\hat{x}$, we then applied a decoder projection \linebreak $W_d \in \mathbb{R}^{N \times K}$ and bias $b_d \in \mathbb{R}^N$:

\begin{equation}
    \hat{x} = W_d f(x) + b_d
\end{equation}

Training minimizes a combination of \emph{fidelity} and \emph{sparsity} losses. The fidelity loss is the mean squared error between the reconstruction and the original vector:

\begin{equation}
    \mathcal{L}_{\text{fidelity}}(x, \hat{x}) = \frac{1}{N}\sum_i^N (\hat{x}_i - x_i)^2 
\end{equation}

We used an updated form of the sparsity loss from \cite{connerly2024april}, but we applied this only to the mask instead of to the combined dictionary activations, as this avoids penalizing the scale of the activations (in the spirit of the $L_0$ loss in \cite{rajamanoharan2024jumping}), while also maintaining the possibility of negative activations:

\begin{equation}
    \mathcal{L}_{\text{sparsity}}(x) = \frac{1}{K}\sum_i^K mask(x)_i \|W_{d, i}\|_2
\end{equation}

With a parameter for loss balancing ($\lambda$), the total SAE loss can be written as:

\begin{equation}
    \mathcal{L}_{\text{SAE}}(x, \hat{x}) = \mathcal{L}_{\text{fidelity}}(x, \hat{x}) + \lambda \cdot \mathcal{L}_{\text{sparsity}}(x)
\end{equation}

\begin{figure}
    \centering
    \includegraphics[width=0.8\linewidth]{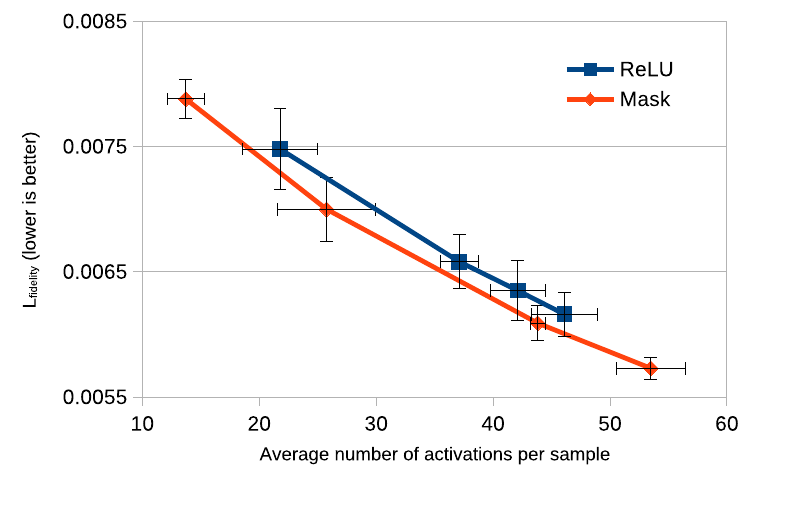}
    \caption{Mask vs. ReLU Activation for four values of $\lambda = 0.03, 0.01, 0.003, 0.001$. Error bars are 95\% confidence intervals computed using Student's t-test over four trials.}
    \label{fig:mask-vs-relu}
\end{figure}

Compared to ReLU-based methods, the mask mechanism allows supports negative activations and separates the activation decision from the computation of entry scale, achieving a better overall sparsity/fidelity trade-off. See Figure~\ref{fig:mask-vs-relu} for a performance comparison of the two activation approaches computed over four values of $\lambda = 0.03, 0.01, 0.003, 0.001$ and four trials.

Once the SAE is trained, we compute correlations between the activations and a large set of hand selected features. A full list of features can be seen in Appendix \ref{apx:features}.

\section{Experiments}
\label{ssec:experiments}

Below, we describe in detail our data, training, and analysis procedure.

\subsection{Parkinson's disease dataset}
\label{ssec:qpn}

For our experiments with Parkinson's disease, we used a set of speech recordings from the Quebec Parkinson Network (QPN) (\cite{gan2020quebec}) -- 208 patients and 52 controls. The demographic breakdown of the study subjects can be seen in Appendix~\ref{apx:breakdown}. A portion of these speech data have previously been used to show that articulation impairments in patients with PD are associated with aberrant activity in the left inferior frontal cortex (\cite{wiesman2023aberrant}). All patients and controls were recorded in the same way with a headset microphone near the mouth and in a quiet room. The majority of patients were recorded in the ON medication state, meaning they had taken their prescribed dopaminergic treatment (e.g., levodopa) prior to the session. Prior work has shown that dopaminergic medication has limited effects on speech production in Parkinson’s disease (\cite{ovallath2017levodopa,cavallieri2021dopaminergic}).

The subjects were asked to perform a number of tasks, such as sustained phonation, reading a passage, recalling a memory, and describing a picture. The latter was selected as initial experiments showed the best discrimination performance -- likely due to the cognitive demands of the task (\cite{bocanegra2015syntax}). The task was introduced as part of the Boston Diagnostic Aphasia Exam (\cite{goodglass1979assessment}), and the picture participants are asked to describe is known as the ``cookie theft'' picture. All tasks were approved by an institutional review board, and the relevant risks were communicated to the participants.

From these data, we selected 32 subjects to form a demographically balanced test set, which contains equal numbers of patients and controls, native French and English speakers, and men and women. To ensure that patients were in the early stages of the disease, we selected subjects with an average time between diagnosis and speech recording of 3.1 years. Finally, we primarily selected patients with mild motor symptoms, as indicated by Unified Parkinson's Disease Rating Scale (UPDRS) II scores of 12 or less and UPDRS III scores of 32 or less (\cite{martin2015parkinson's}).
For the training set, the data are not balanced. We use sampling and weighting strategies to address the patient status and sex imbalances. 

As for language, although native French speakers outnumber other speakers by a ratio of 2.2 to 1, 59\% of them also speak English and for these subjects we have recordings of tasks in both languages. This makes the final language distribution more balanced.

\subsection{Speech-based detector training}
\label{ssec:experiment_detector}

We experimented with a minimal configuration of parameters on top of a set of frozen encoders. For each encoder, we train a small classifier consisting of the following layers: linear, attention pooling across time, linear, output (binary). Between linear layers, we added dropout at a rate of 0.2 and leaky ReLU activations.

We trained all models with an Adam 8-bit optimizer and used a learning rate of 0.0001 with linear warm-up for 2 epochs and cosine cool-down for the remaining 18 epochs. Every epoch consists of 1024 samples with 32 samples per batch.

We used sampling and weighting techniques to address the data imbalances in terms of sex and subject type. For all experiments, we equally sample from the four categories created by the intersection of these two conditions. In addition to adjusting the sampling for sex and patient status, we also re-weighted samples, at 0.7 for majority types and 1.5 for minority types. We sample 1024 utterance chunks of 30 seconds or less each epoch.

To reduce the incidence of overfitting on the over-sampled data, we introduced data augmentations on the waveforms of 90\% of batches. We combined two augmentations: adding background noises at random signal-to-noise ratios between 0 and 15 and dropping between 2 and 5 random frequencies via notch filters.

In addition to pretrained models, we provide results on mel filterbank and extracted vocal features, which serve as baselines for comparing the pretrained model features. The vocal features are a combination of autocorrelation-based features (f0, harmonicity, glottal-to-noise-excitation ratio, see \cite{godinollorente2010gne}), period-based features (jitter, shimmer), spectrum statistical features (e.g. skew, flux), and the first 4 mel-frequency cepstral coefficients.

\subsection{Training SAEs}
\label{ssec:experiment_sae}

Once the detection model is trained, we insert the SAE layer immediately after the attention pooling, so that one dictionary vector is created for the whole sample, rather than one per frame. We use 64 entries for the size of our dictionary. While typical SAEs use dictionaries up to sizes much larger than the size of the explained layer, for our small datasets and our choice of applying the SAE after the attention pooling, we do not see any benefit from larger dictionaries as there are relatively few explanations needed. We trained the SAE with a relatively high learning rate of 0.003 using an Adam optimizer to improve sparsity, and used a sparsity weight of $\lambda=0.001$.

The final fidelity score for our system on the test data is 0.0068 which indicates a close match with the true activations. We also verify that the system has a high fidelity by computing the predictions by replacing the original attention pooling outputs with the predicted pooling outputs from the SAE layer, and find a less that 2\% absolute drop in F1 score, indicating that the system is operating in much the same way, even with the lossy compression from the SAE layer.

\section{Results}
\label{sec:results}

\begin{figure}
    \centering
    \includegraphics[width=0.4\linewidth]{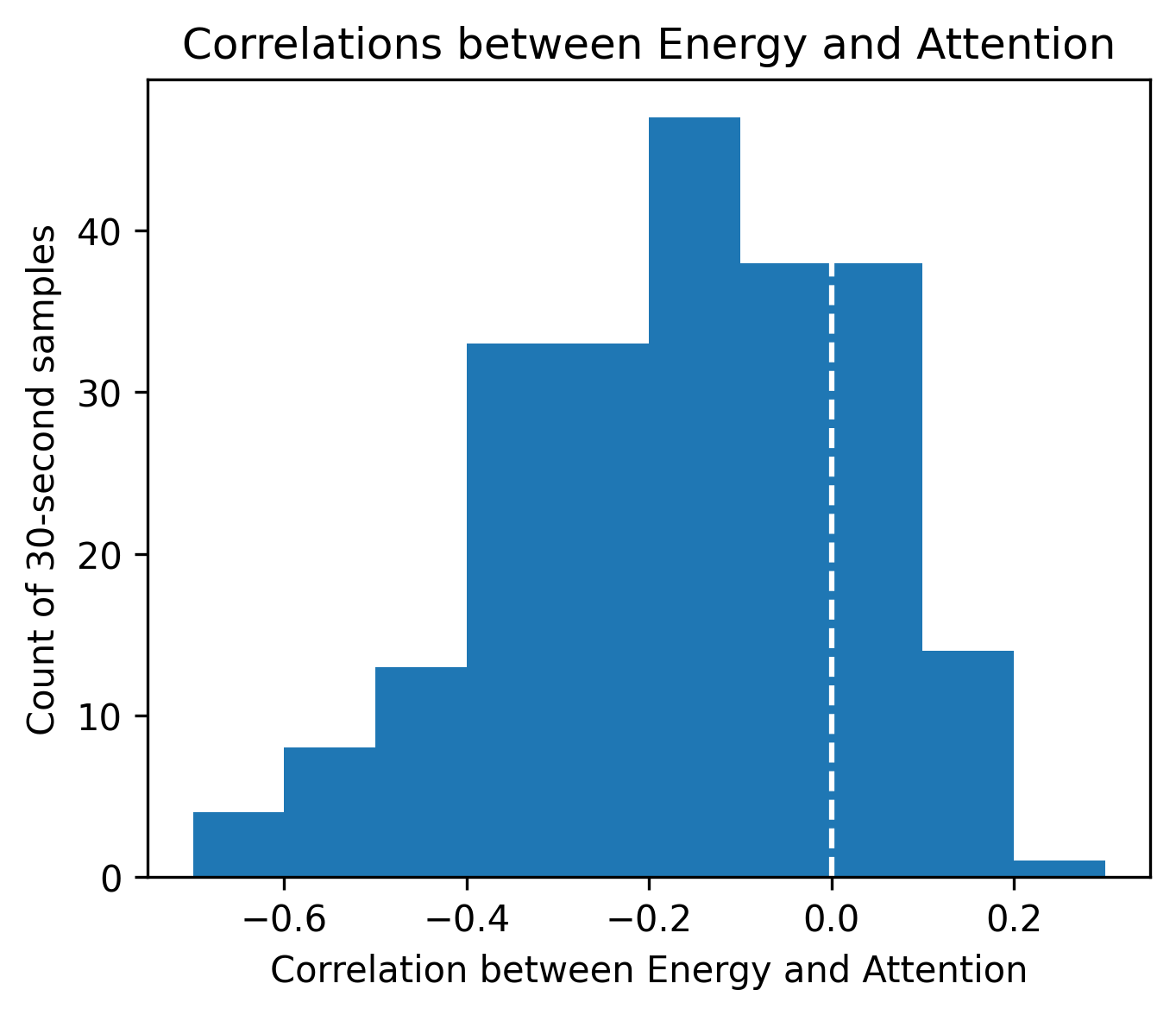}
    \includegraphics[width=0.59\linewidth]{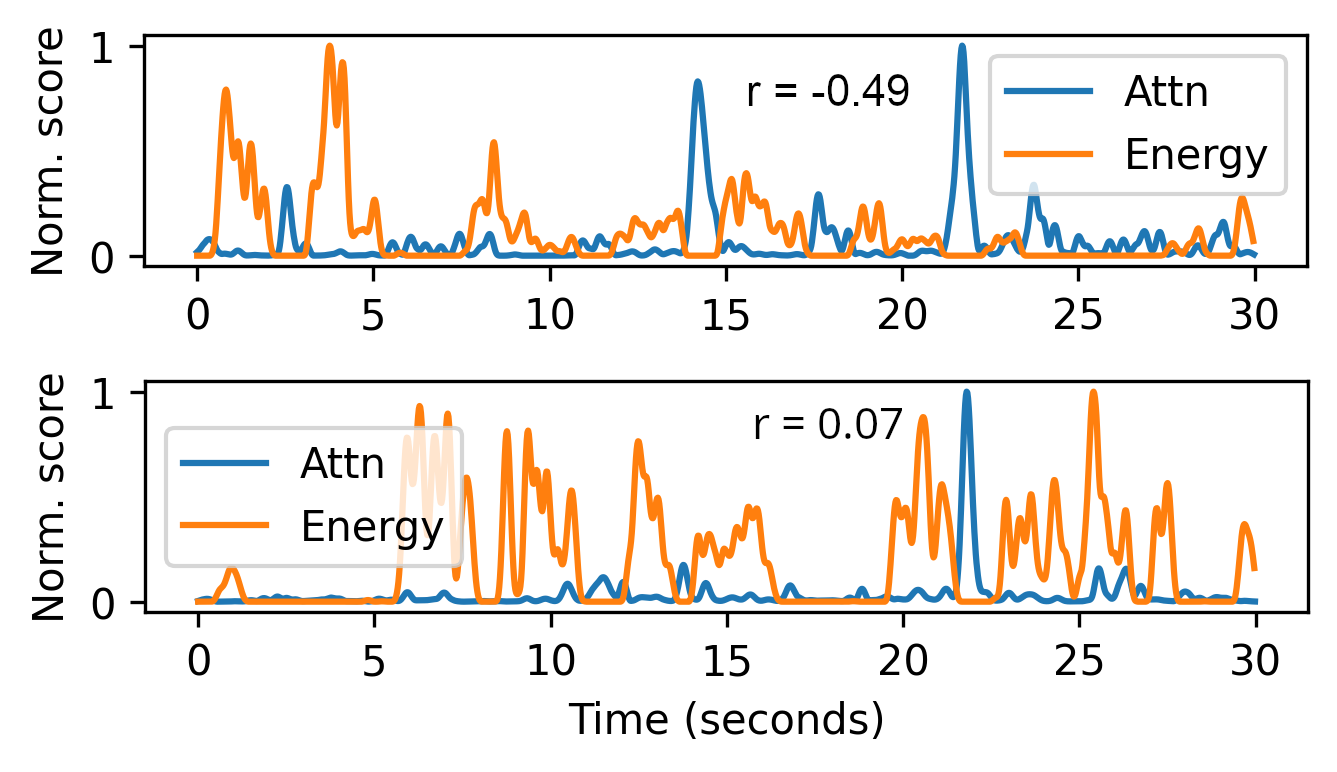}
    \caption{Attention shows a pattern of anticorrelation with the signal energy over the test set (left). Correlation is computed by smoothing the signal, then binarizing the attention and energy signals, > 5\% of max is "active" and the rest is "inactive", and then computing the cross-correlation. Individual samples are shown on the right.}
    \label{fig:energy_anticorrelated}
\end{figure}

The encoder comparison for the task of Parkinson's disease detection, seen in Appendix \ref{apx:detection}, validate that our system is finding important features of the disease for early detection. Our best performing systems achieve better than 80\% F1 score overall.

\subsection{Sparse Autoencoder}
\label{ssec:results_sae}

With these results in place, we train a sparse autoencoder on the Whisper Small system. Our first investigation in interpretability uses the pattern of attention to localize the activity of the network in the feature embedding. When plotted against the energy of the signal, as in Figure \ref{fig:energy_anticorrelated}, we see a pattern of anticorrelation, meaning the network tends to pay attention to areas of low energy. This means the regions have no speech or the attended speech seems to be weak or breathy. Although initially unexpected, we have found other work that notices similar correlations. \cite{darling2020impact} find that breath pauses contains signs of Parkinson's disease. But its not just pauses, but also the breath itself that can indicate Parkinson's disease, \cite{yang2022artificial} found that nightly breathing patterns can reveal signs of Parkinson's disease.

\subsection{Correlation and statistical significance}
\label{ssec:correlation}

In addition to the pattern of attention, we analyze correlations between the activations of the SAE and the set of interpretable vocal features provided by SpeechBrain version 1.0.2 (\cite{speechbrainV1}). We find that strong correlations are present with features that can be interpreted with respect to known features of the speech of people with Parkinson's disease, the result of which can be seen in Table \ref{tab:entry_correlations}. The strongest correlation is with the spectral flux, which is related to the strength of the speaker's articulation. Healthy controls tend to have more dynamic speech, which will have a larger spectral flux, while patients are known to have flatter articulatory trajectories (\cite{schulz2000effects}). The spectral flux $\Phi$ is computed as the mean-square-difference of each successive pair of spectral frames $S_t, S_{t+1}$, which we then weight according to the attention scores $A_t$ given by our pooling layer.

\begin{equation}
    \Phi = \frac{1}{T} \sum_{t=0}^T \frac{A_t}{F} \sum_{f=0}^F (S_{t+1,f} - S_{t,f})^2
\end{equation}

Another key feature used for the prediction seems to be the spectral flatness, which is one way to measure the degree of noise in phonations. Parkinson's disease does sometimes cause breathy or raspy phonations, due to reduced control of the glottal folds (\cite{schulz2000effects}).

One interesting insight is that the network does seem to attend to the language of the recording, but does not base the final prediction score on the language. Perhaps there are some differences in vowel pronunciation between the language, and the language can be a moderating effect on the other scores.

There are also notable absences from the table of strongest correlations: there were no strong correlations for other factors that could be confounding or predictive, such as the sex, age, number/length of pauses, or specific word types. While we can't comprehensively rule out the possibility that some of these variables are represented in a distributed way across the dictionary entries, it is encouraging that language emerges as one important factor, but the other confounding factors do not show any strong correlations.

\begin{table}
    \centering
    \caption{List of the dictionary entries most correlated (using Spearman's rank) with interpretable features and the corresponding correlation to the final prediction.}
    \label{tab:entry_correlations}
    \begin{tabular}{cclcc}
        \toprule
        Entry & Count & Interpretable feature & Activation Corr. & Prediction Corr.  \\
        \midrule
        \#06 & 229 & Harmonic-to-Noise ratio (phonation) & $\rho$ = -0.682 & $\rho$ = 0.794 \\
        \#13 & 129 & Spoken Language (En or Fr) & $\rho$ = 0.738 & $\rho$ = -0.239 \\
        \#26 & 229 & Spectral flatness (phonation) & $\rho$ = 0.816 & $\rho$ = 0.975 \\
        \#61 & 229 & Spectral flux (articulation) & $\rho$ = 0.851 & $\rho$ = -0.943 \\
        \bottomrule
    \end{tabular}
\end{table}

\begin{figure}
  \centering
  \includegraphics[width=0.7\linewidth]{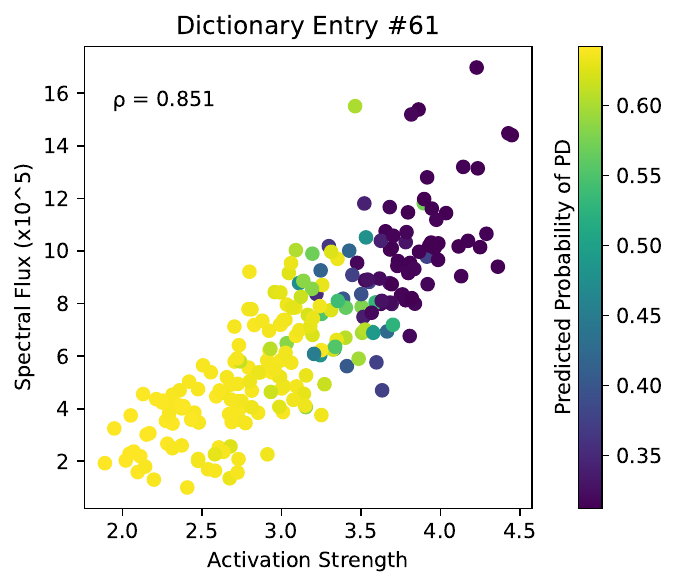}
  \caption{Correlation of spectral flux with dictionary entry \#61 and network prediction.}
  \label{fig:spectral_flux_correlation}
\end{figure}

Figure \ref{fig:spectral_flux_correlation} plots the activation strength of dictionary entry \#61 against the spectral flux of the attended region. The correlation is highly statistically significant. We computed the significance by randomly selecting one sample from each speaker (total $N=32$) to avoid artificially inflated correlations and for a conservative estimate we applied the Bonferroni adjustment for the number of correlation tests performed (33 features $\times$ 64 dictionary entries = 2112 tests). The result remained highly statistically significant ($p < 10^{-20}$) indicating that this entry reliably captures spectral flux variations.

\subsection{Brain region volumetric measurement correlations}
\label{ssec:mri}

\begin{figure}
  \centering
  
  \includegraphics[height=140pt]{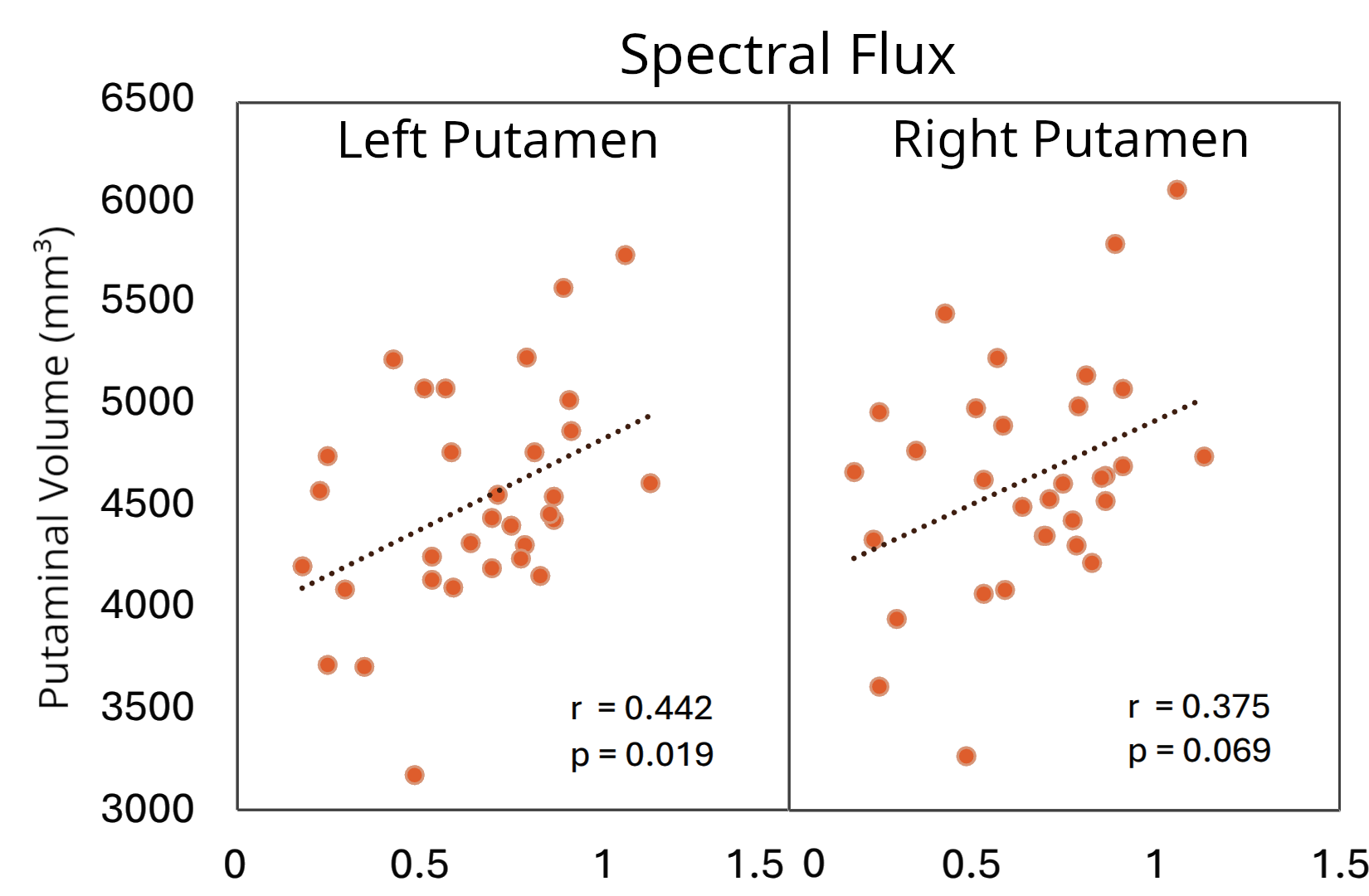}
  \hfill
  \includegraphics[height=140pt]{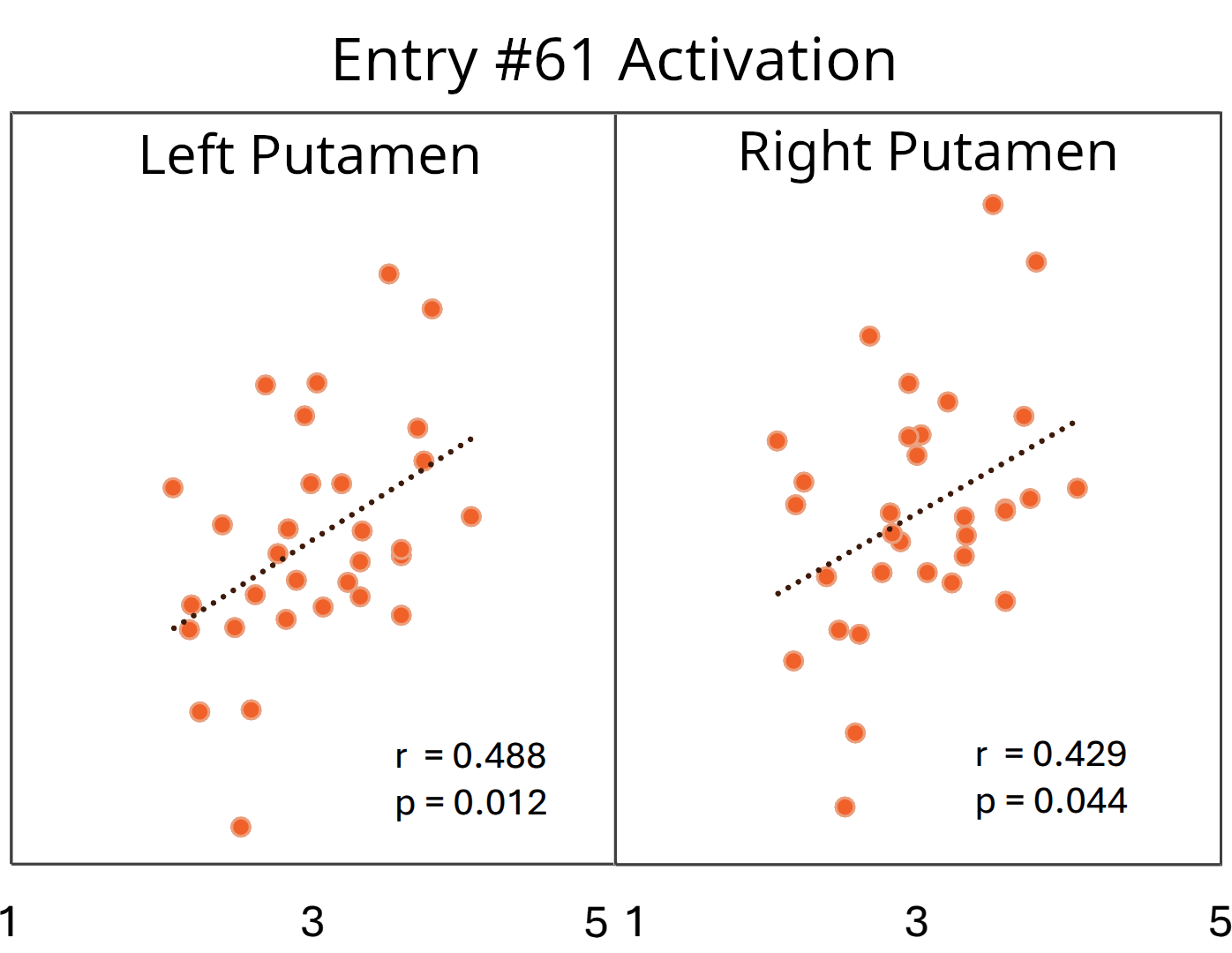}
  \caption{Correlation of spectral flux (left) and the activation strength of entry \#61 (right) with the volume (in $mm^3$) of the left and right putamen.}
  \label{fig:putamen}
\end{figure}

Finally, we check the correlation of the spectral flux and the activation strength of entry \#61 with the volumetric measurement of the different sub-regions of the basal ganglia in a MRI scan taken at a similar time as the speech recordings. We tested correlations with four sub-regions in the basal ganglia: caudate nucleus, putamen, pallidum, and Accumbens area. As the spectral flux and activation strength can vary across samples for a given participant, we average the results from each sample to achieve a single score per participant. We see a significant correlation for the putamen, but not the other sub-regions, indicating that the dictionary entry may be revealing an important factor of the brain's behavior. The correlation can be seen in Figure \ref{fig:putamen}.

The putamen is known to play a role in motor control, especially for language (\cite{ghandili2023neuroanatomy}). In addition, early stages of Parkinson's disease has been shown to be associated with signs of atrophy in the putamen by \cite{kinoshita2022putamen}. 



\section{Limitations}
\label{sec:limitations}

While our approach demonstrates a promising new direction for speech biomarker design, there are some potential limitations to be aware of in our research. First, while Parkinson's disease may be a good condition for testing potential speech biomarkers due to its well-known effects on speech, it remains unclear how well the SAE technique would generalize to other neurological conditions. It is possible that this technique would have limited correlations with known effects of a given condition on patient speech. Second, and related to the first, this approach requires knowledge of the effects of a tested condition on speech in order to know what sorts of correlations to look for. The space of possible correlations can only be so large before spurious correlations become likely. Third, the interpretations of the SAE are necessarily indirect, and may miss patterns that are meaningful. Finally, our novel mask-based SAE approach has not been fully validated as to its effectiveness on a wider variety of architectures and problems. More extensive comparisons with other approaches and on other datasets would help to validate its utility.

\section{Conclusions}
\label{sec:conclusions}

In this work, we demonstrate that sparse autoencoders (SAEs), when adapted to the constraints of small biomedical datasets, can effectively uncover interpretable internal representations in deep learning models for speech-based Parkinson’s disease detection. Our results show that attention mechanisms and sparse representations together provide valuable insights into model behavior. Specifically, we find that the model bases its predictions on acoustic properties such as spectral flux and spectral flatness—features that align with known articulatory deficits in Parkinsonian speech. Furthermore, we link these representations to structural brain changes by demonstrating a correlation between spectral flux and putamen volume in MRI scans, providing evidence that the model’s focus corresponds to meaningful neuroanatomical variation.

These findings support the potential of SAEs to enhance transparency in deep learning models, and to advance biomarker discovery. As machine learning becomes increasingly integrated into clinical decision-making, approaches like ours can start to build trust with clinicians. Future work should explore the application of this method to a wider range of disorders and assess whether these insights can support longitudinal tracking or treatment planning.


\bibliographystyle{plainnat}
\bibliography{refs}





\newpage
\begin{appendices}

\section{Quebec Parkinson's Network (QPN) dataset demographics}
\label{apx:breakdown}

\begin{table}[h]
    \centering    
    \begin{tabular}{ll}
        \toprule
        \multicolumn{2}{c}{\textbf{208 Patients}} \\
        \midrule
        Sex & Male 133 -- Female 75 \\
        Age & mean 65.9 yr. -- standard deviation 8.7 yr. \\
        First language & French 73\% -- English 13\% -- Other 14\% \\
        UPDRS III & mean 29.6 -- standard deviation 13.6 \\
        Years since diagnosis & mean 4.1 \\
        \midrule
        \multicolumn{2}{c}{\textbf{52 Controls}} \\
        \midrule
        Sex & Male 16 -- Female 36 \\
        Age & mean 63.7 yr. -- standard deviation 8.8 yr. \\
        First language & French 44\% -- English 34\% -- Other 22\% \\
        \bottomrule
    \end{tabular}
\end{table}

The above table consists of a list of some basic demographic statistics of the QPN speech dataset. The demographic data show that these samples are useful for detection, as the average number of years since diagnoisis is less than 5 years, and the average UPDRS III score is less than 32, which is considered the cut-off for "mild" motor symptoms of disease. In addition, the age distribution is balanced between the controls and PD patients. The sex and language are not balanced in the training set, but we carefully select samples for our test set that balance age, sex, language, and patient status (PD / HC), and use sampling and loss weighting and augmentation strategies to adapt the training dataset for more balanced model training and better performance on the balanced test set.

\section{Encoder comparison results}
\label{apx:detection}

\begin{table}[h]

  \centering
  \begin{tabular}{lccccc}
    \toprule
    \textbf{Encoder}     & \textbf{Params} & \textbf{Fr F1} & \textbf{En F1} & \textbf{F1} \\
    \midrule
    SB VocalFeats        & 0.6M & 74.7 & 59.0 & 65.9 \\
    Filterbank n=80      & 0.7M & 66.2 & 54.0 & 60.0 \\
    OS eGeMAPSv02        & 0.6M & 52.6 & 54.6 & 53.7 \\
    \midrule
    Whisper Small        & 89M  & \textbf{87.5} & 75.0 & \textbf{81.1} \\
    Whisper Medium       & 309M & 82.4 & \textbf{82.4} & \textbf{82.4} \\
    XEUS                 & 579M & 84.2 & 70.0 & 78.9 \\
    WavLM Base+          & 96M  & 84.2 & 70.6 & 76.5 \\
    WavLM Large          & 318M & 80.0 & 70.6 & 75.7 \\
    HuBERT Base          & 96M  & 80.0 & 66.7 & 73.7 \\
    Whisper Large v3     & 639M & 80.0 & 63.2 & 71.8 \\
    wav2vec 2.0 Base     & 96M  & 78.7 & 56.4 & 68.8 \\
    Crisper Whisper      & 639M & 73.7 & 59.4 & 68.5 \\
    Whisper Small.en     & 89M  & 72.7 & 66.7 & 68.3 \\
    Parakeet CTC 0.6B    & 610M & 72.7 & 53.3 & 64.9 \\
    \bottomrule
  \end{tabular}
\end{table}

Comparison of encoders for use in a speech biomarker system for monitoring of Parkinson's disease. Above the midline are shallow encoders with interpretable features, and below the midline are deep encoders that extract more predictive--but less interpretable--features. We present the median score over five runs. We note that the results presented here are not for the purpose of achieving the best-performing system but rather to find a good-performing system that we can further analyze for interpretability. Another note is that the interpretable features used for explanation are not by themselves accurate enough for use as a speech biomarker system.

\pagebreak

\section{Set of interpretable features used for SAE correlation analysis for Parkinson's disease detection}
\label{apx:features}

\begin{table}[h]
    \centering
    \begin{tabularx}{\textwidth}{lXl}
        \toprule
        Category                & Feature list & Best correlation \\
        \midrule
        Demographics & Sex, Age, Sample language (French or English), First language (French, English, Other), Patient status (PD / HC) & Language -> 0.77 \\
        Period-based statistics & fundamental frequency (F0) estimated via autocorrelation, harmonic-to-noise ratio (HNR), jitter, shimmer, glottal-to-noise excitation ratio (GNE) & HNR -> 0.68 \\
        Spectral statistics     & spectral centroid, spectral spread, spectral skew, spectral kurtosis, spectral entropy, spectral flatness, spectral crest, spectral flux, MFCC 1st-4th coefficients & spectral flux -> 0.85 \\
        Pause statistics        & length of longest pause, count of pauses longer than one second, total length of non-speech sections, ratio of non-speech to speech & speech ratio -> 0.41 \\
        Lexical statistics      & word count, word count per second, sentence count, sentence count per second, average sentence length, average word length & word count -> 0.44 \\

        \bottomrule
        
    \end{tabularx}
\end{table}

The above table contains the full set of features we use to test our system, selected out of a variety of categories so as to accurately determine what type of category may be the basis for the model decision. In addition, we list the Spearman's rank correlation of the most correlated feature as an estimation of the quantitative role that each category plays in the final decision.





\end{appendices}

\end{document}